\pdfoutput=1
\documentclass[a4paper,
               ]{jacow}

\usepackage{pdfpages,multirow,ragged2e} %

\makeatletter%
	\ifboolexpr{bool{xetex}}
	 {\renewcommand{\Gin@extensions}{.pdf,%
	                    .png,.jpg,.bmp,.pict,.tif,.psd,.mac,.sga,.tga,.gif,%
	                    .eps,.ps,%
	                    }}{}
\makeatother

\ifboolexpr{bool{xetex} or bool{luatex}} % test for XeTeX/LuaTeX
 {}                                      % input encoding is utf8 by default

\usepackage[USenglish]{babel}
\usepackage{hyperref}
\usepackage{cleveref}
\usepackage{graphicx}
\DeclareGraphicsExtensions{.pdf,.png,.jpg}
\usepackage{subcaption}
\usepackage{listings}
\begin{document}

\title{application of large language models for the extraction of information from particle accelerator technical documentation}

\author{Q. Dai\thanks{qing.dai@psi.ch}, R. Ischebeck, M. Sapinski, Paul Scherrer Institut, Villigen, Switzerland\\
A. Grycner, Google DeepMind}
% add EPFL
    
\maketitle

\begin{abstract}
 The large set of technical documentation of legacy accelerator systems, coupled with the retirement of experienced personnel, underscores the urgent need for efficient methods to preserve and transfer specialized knowledge. This paper explores the application of large language models (LLMs), to automate and enhance the extraction of information from particle accelerator technical documents. By exploiting LLMs, we aim to address the challenges of knowledge retention, enabling the retrieval of domain expertise embedded in legacy documentation.
We present initial results of adapting LLMs to this specialized domain. Our evaluation demonstrates the effectiveness of LLMs in extracting, summarizing, and organizing knowledge, significantly reducing the risk of losing valuable insights as personnel retire. Furthermore, we discuss the limitations of current LLMs, such as interpretability and handling of rare domain-specific terms, and propose strategies for improvement. This work highlights the potential of LLMs to play a pivotal role in preserving institutional knowledge and ensuring continuity in highly specialized fields.
\end{abstract}

\maketitle

%%%%%%%%%%%%%%%%%%%%%%%%%%%%%%%%
\section{Introduction}

Having a well-structured and complete documentation is crucial for maintaining and developing any large industrial system.
In case of accelerator facilities, such as the High Intensity Proton Accelerator (HIPA) \cite{Seidel:2010zz} at PSI, the available documentation is often sparse and inaccurate. 
HIPA has been designed and build 50 years ago, before any electronic documentation was invented. 
Furthermore, as an experimental facility, it evolved significantly over years. 
The absence of a consistent long-term documentation strategy has further contributed to the current state of confusion.

Many accelerators facilities face similar problems. Newer machines usually have more consistent documentation. 
For instance Proscan, which is a proton therapy facility in PSI, 
is well documented as it was required by licensing authorities. 
Also any change done to medical facility must be carefully analyzed and approved before
implementation. As a result Proscan, which is 20 years old, is almost immutable with respect to HIPA.

The expertise is not only in bare set of documentation.
% not exhaustive,  on-hand expertise etc etc
When a new specialist takes over the responsibility, he or she needs a certain amount of time, counted in years, to reach the same level of performance as the retiring expert. % Overlap of the generations is crucial for transferring the knowledge.
Simulating the ability to chat (questions and answers) with an expert can be a great tool to speed up 
this process.

In this work we took beam instrumentation of HIPA and Proscan as an example to develop LLM-based chatbot which could help, based on existing documentation, to work on technical issues of the facility. Similarly to other labs \cite{Mayet:2024ypj,Sulc:2024ssg}, we choose to develop a system which runs locally, not exposing internal documentation.

%%%%%%%%%%%%%%%%%%%%%%%%%%%%%%%%
\section{The documents} \label{sec:the documents}

The initial set of documents included 58 pdf files, in English (36) 
and in German (22), written mostly by one physicist over the span of 30 years. 
Those documents are technical and specific to HIPA and Proscan.
Before using them, they were checked if they are up to date.
In addition a few other documents were included:
\begin{itemize}
    \item Two master theses, containing general descriptions of HIPA and results of particular investigations.
    \item Several conference proceedings which include overview of HIPA and Proscan instrumentation \cite{Dolling:2003yw} or more detailed analysis of problems related to particular devices, for instance \cite{Sapinski:2022aqv}.
    \item Publicly available books with courses on beam instrumentation eg.\cite{Forck2011}.
\end{itemize}

During the tests it was found that an important part of the information - for instance the naming schemes or the location of electronics racks - is missing. Large part of these information are present in excel tables, schematics and databases, which we cannot process yet. The gap was partially filled by creating dedicated text files with the missing information. 

The input documents are stored in \textit{Corpus} directory and a script to process them is provided.
In total they contain 1032 pages in English and 239 pages in German.

%%%%%%%%%%%%%%%%%%%%%%%%%%%%%%%%

\section{Methods}
\label{sec:methods}

Retrieval-augmented generation (RAG) combines the broad language ability of large language models (LLMs) with an external knowledge base, improving factual accuracy and grounding~\cite{gao2023retrieval}.  
%In our study the knowledge base is the accelerator-domain document set described in Section~\ref{sec:the documents}.  
All data were processed offline, and every model was run locally with \textbf{Ollama} \cite{ollama} on a Mac Studio M2 Ultra (192 GB unified memory).

\begin{figure}[htb]
  \centering
  \includegraphics[width=\linewidth]{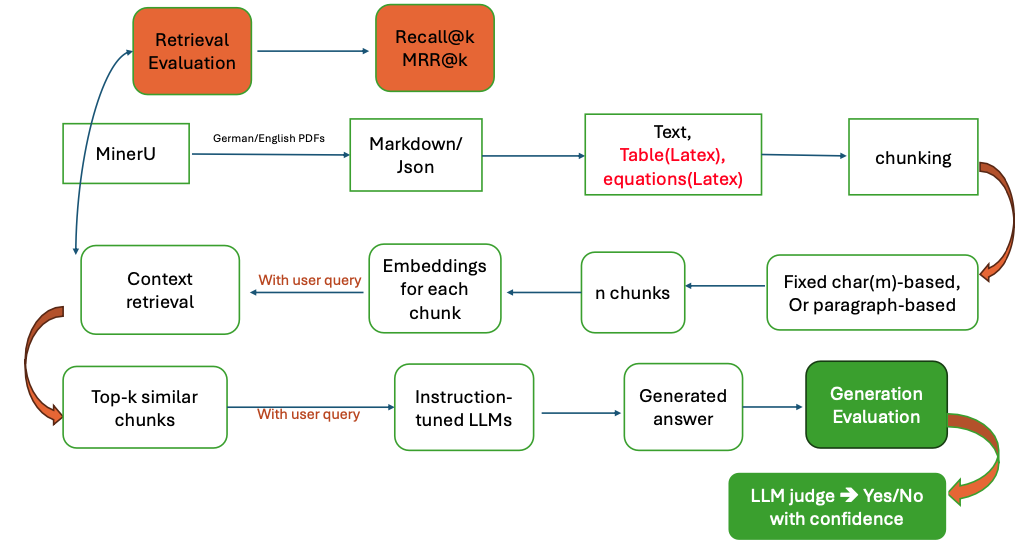}
  \caption{End-to-end RAG pipeline.}
  \label{fig:pipeline}
\end{figure}

\textbf{Pipeline.}  
Figure~\ref{fig:pipeline} shows the two-stage RAG workflow:

\begin{enumerate}[leftmargin=*,itemsep=2pt]
  \item \textbf{Pre-processing.}  
        PDFs are parsed with the open-source \textsc{MinerU} extractor~\cite{Wang:2024mineru}.  
        We retain text, equations, and tables, then split each document into chunks and store
        (chunk $\to$ embedding, file-id) tuples in a vector database.  Embeddings are pre-computed for fast retrieval.
  \item \textbf{Runtime.}
    \begin{enumerate}[label=(\alph*),itemsep=2pt]
      \item \textit{Retrieval.}  
            A user query is embedded with the same multilingual BGE-M3 model~\cite{Chen:2024bge}; the top-$k$ similar chunks are returned.
      \item \textit{Generation.}  
            The query plus retrieved chunks are fed to the instruction-tuned \texttt{gemma3:27b-it-fp16} \cite{GemmaTeam2025Gemma3} LLM, which produces the final answer.
      \item \textit{Evaluation.}  
            Retrieval is scored with recall@k and MRR@k~\cite{Pinecone:2023}; generation is scored with answer accuracy given a reference answer and mean confidence over correctly judged answers, using the same Gemma model as a judge.
    \end{enumerate}
\end{enumerate}

\begin{figure}[htb]
  \centering
  \begin{subfigure}[b]{\columnwidth}
    \centering
    \includegraphics[width=\linewidth]{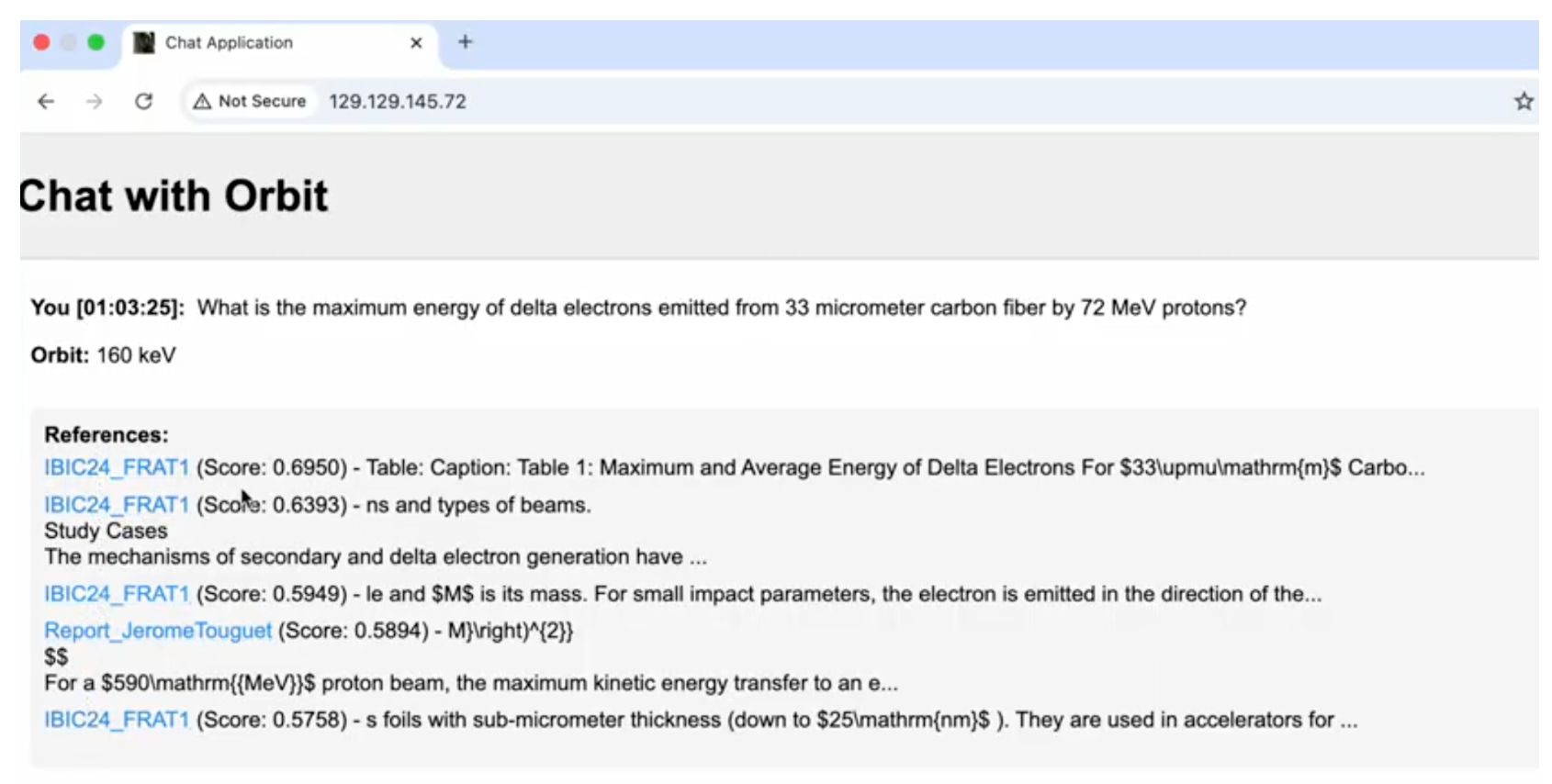}    
    \caption{Query and LLM answer.}
    \label{fig:chatbot1}
  \end{subfigure}\\[0.8ex]  % tiny vertical gap
  \begin{subfigure}[b]{\columnwidth}
    \centering
    \includegraphics[width=\linewidth]{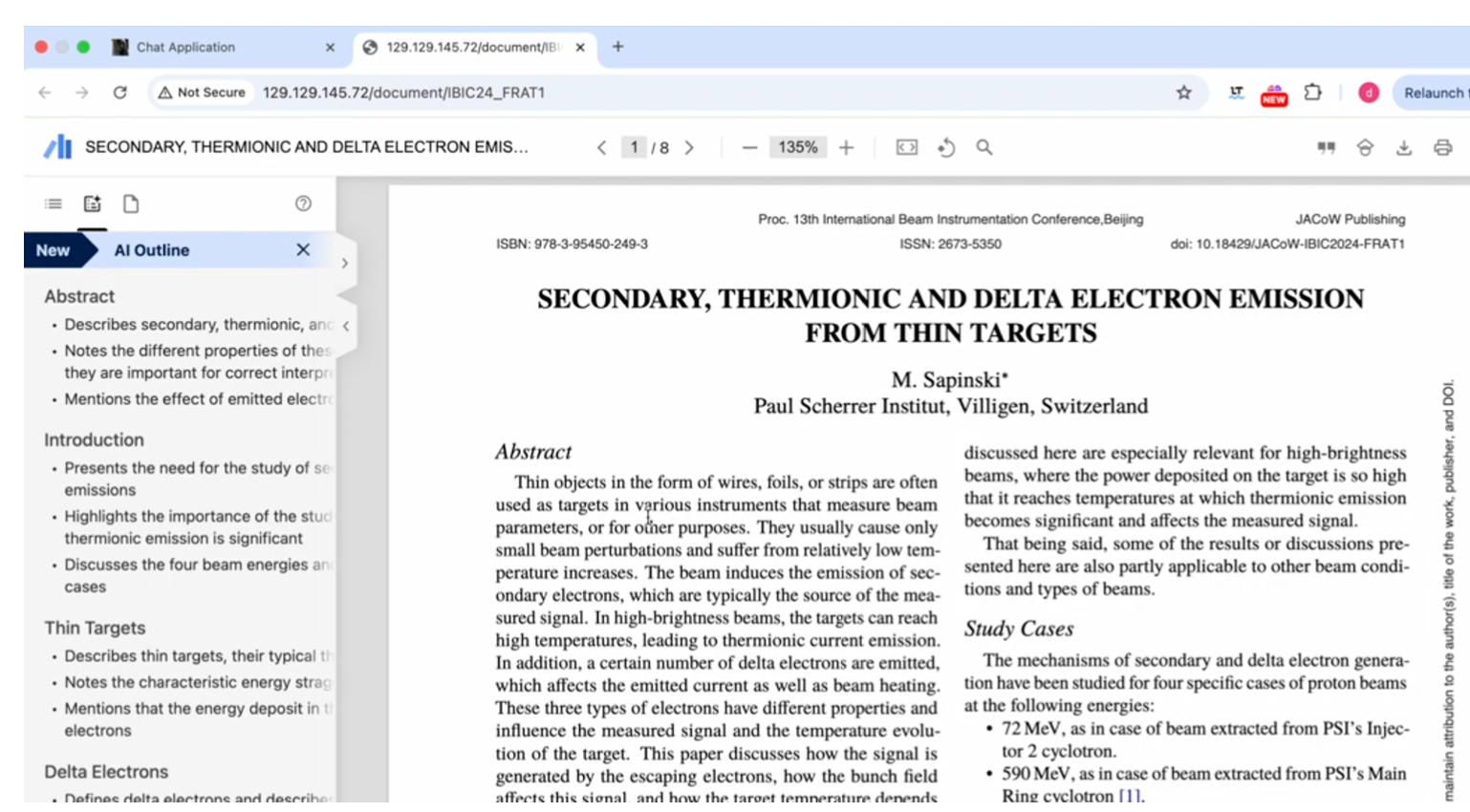}    
    \caption{Clickable document reference.}
    \label{fig:chatbot2}
  \end{subfigure}
  \caption{Chatbot interface.}
  \label{fig:chatbot}
\end{figure}

\textbf{Chatbot interface.}  
Figure \ref{fig:chatbot} illustrate the front-end. 
Besides the answer, the bot lists the five most relevant files with
similarity scores and snippets; clicking a file opens the exact location in the context.
%, saving users from manually skimming hundreds of pages.

\textbf{Benchmark.}  
%To our knowledge this is the first RAG study focused on accelerator documentation.  
Two domain experts created 100 question–answer (QA) pairs, each linked to a gold reference file (70 English, 30 German).  
We use this set for both retrieval and generation evaluation. 
%detailed results appear in Section~\ref{sec:results}.

% -------------------------------------------------------

\section{Results}
\label{sec:results}

\subsection{Chunking strategies}
We tested four splitting schemes:

\begin{itemize}[leftmargin=*,itemsep=2pt]
  \item character windows of 800, 1600, and 2000 chars;
  \item paragraph windows (minimum 120 tokens, with short paragraphs merged).
  \item paragraph with context (previous and following paragraph): this scheme is used in generation, hoping to provide more context from offset, due to semantics which may be broken by chunking.
\end{itemize}

\subsection{Retrieval performance}

\begin{figure}[ht]
  \centering
  \begin{subfigure}[b]{\columnwidth}
    \includegraphics[width=\linewidth]{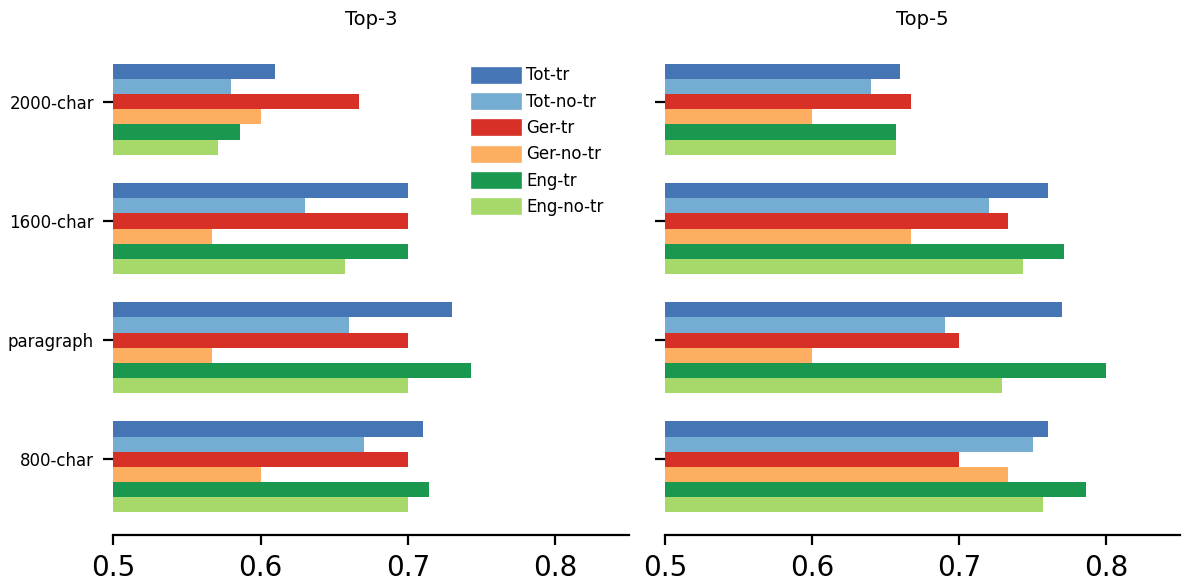}
    \caption{Recall@k.}
  \end{subfigure}\\[1ex]
  \begin{subfigure}[b]{\columnwidth}
    \includegraphics[width=\linewidth]{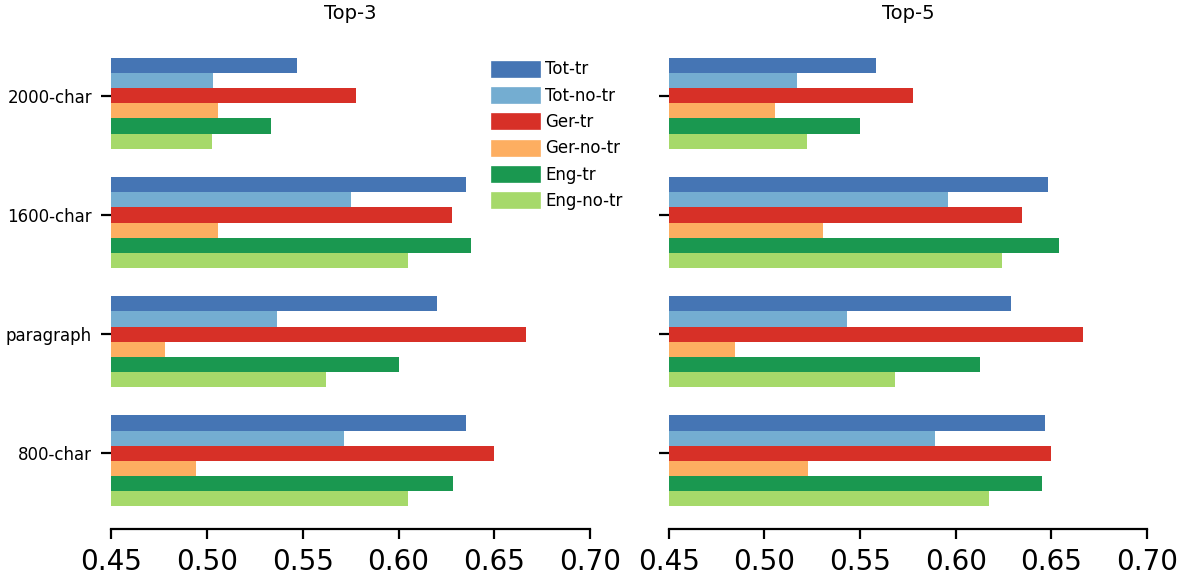}  
    \caption{MRR@k.}
  \end{subfigure}
  \caption{Retrieval comparison across chunking schemes before and after translation.}
  \label{fig:retrieval}
\end{figure}

Figure~\ref{fig:retrieval} summarizes recall@k and MRR@k for $k\!\in\!\{3,5\}$ across all chunk sizes.  
Key observations:

\begin{enumerate}[leftmargin=*,itemsep=2pt]
  \item \textbf{Top-5 $\!>\!$ Top-3 for recall}, but MRR grows more modestly, reflecting the trade-off between depth and ranking quality.
  \item \textbf{Smaller chunks outperform larger ones.}  
        Neither 1600- nor 2000-char windows improved recall or MRR. % relative to 800-char.
  \item \textbf{Paragraph splitting offered no clear gain}, despite providing semantically complete units.
  \item \textbf{German queries lagged behind English}.  
        Translating German chunks to English—using the 4-bit \texttt{gemma2:27b-instruct-q4\_K\_M} \cite{GemmaTeam2024Gemma} model—significantly boosted recall and MRR for German queries and lifted English queries, likely by reducing multilingual noise.
\end{enumerate}

%%%%%%%%%%%%%%%%%%%%%%%%%%%%%%%%%%%%%%%%%%%%%%%%%
\subsection{Generation performance}

Guided by retrieval results, we tested generation on the best two chunk sizes (800 and 1600 chars) and paragraph schemes with $k\!\in\!\{3,5\}$ and three prompt variants: \emph{no-translation} (k-N), \emph{translation} (k-T) and \emph{translation + chunk-score} (k-S) (the similarity score is appended and the LLM is instructed to focus on high-score chunks).
Figure~\ref{fig:generation} reports answer accuracy and mean confidence. Key findings are:

\begin{figure}[ht]
  \centering
  \begin{subfigure}[b]{\columnwidth}
    \includegraphics[width=\linewidth]{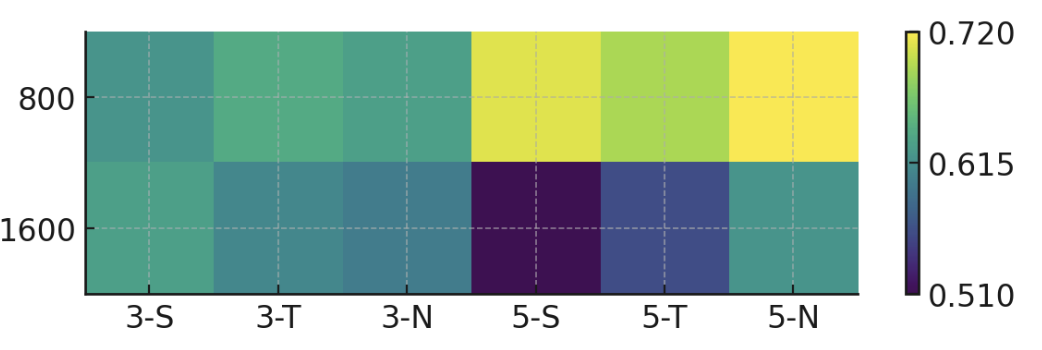}
    \caption{Answer accuracy.}
  \end{subfigure}\\[1ex]
  \begin{subfigure}[b]{\columnwidth}
    \includegraphics[width=\linewidth]{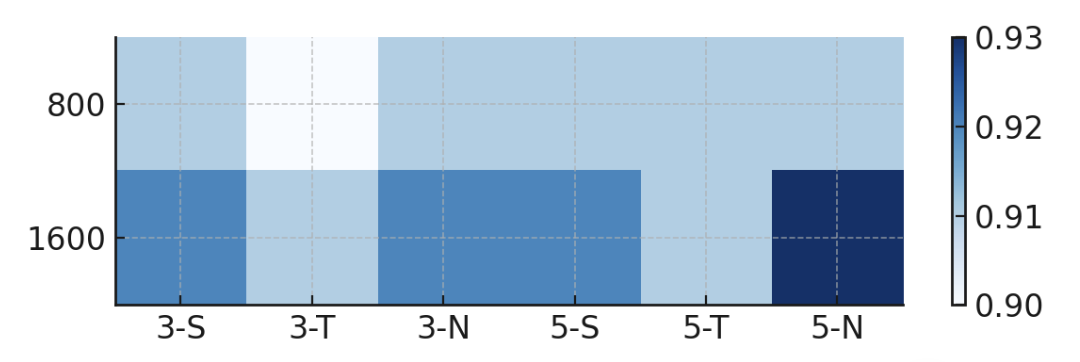}
    \caption{Mean confidence.}
  \end{subfigure}
  \caption{Generation results for 800- and 1600-char chunks under three prompt settings. 
  %k-S: translated chunks appended with scores, k-T: chunks after translation, k-N: chunks without translation
  }
  \label{fig:generation}
\end{figure}

%\textbf{Findings.}

\begin{itemize}[leftmargin=*,itemsep=2pt]
  \item \textbf{Translation helps at $k\!=\!3$ but not consistently at $k\!=\!5$}.  
        Noise from additional chunks partly offsets the translation gain.
\item Overall, the model judges the generated answer with high confidence, with a narrow range from 0.90 to 0.93.

\item While Top-5 retrieval with 1600-character chunks matches the recall of 800-character chunks, it yields lower answer accuracy.  Inspection of the misclassified outputs shows that only this setup suffers from hallucinations—either summarizing retrieved text out of context or ignoring the query altogether—errors absent in the other three chunk–$k$ combinations.  We traced the issue to Ollama’s default context window of 2048 tokens~\cite{Ollama_context_window_2024}.
%(expanded to 4 096 in 2025~\cite{Ollama_context_window_2025}).  
Although our Top-5 1600-char inputs average around 1500 tokens, they approach the limit, leading to truncation and hallucinations.  

\end{itemize}

%%%%%%%%%%%%%%%%%%%%%%%%%%%%%%%%%%%%%%%%%%%%%%%%%%%%%%%%%%%%
\subsection{Paragraph‐Level Chunking and Context Window}

\begin{figure}[htb]
  \centering
   \includegraphics[width=\linewidth]{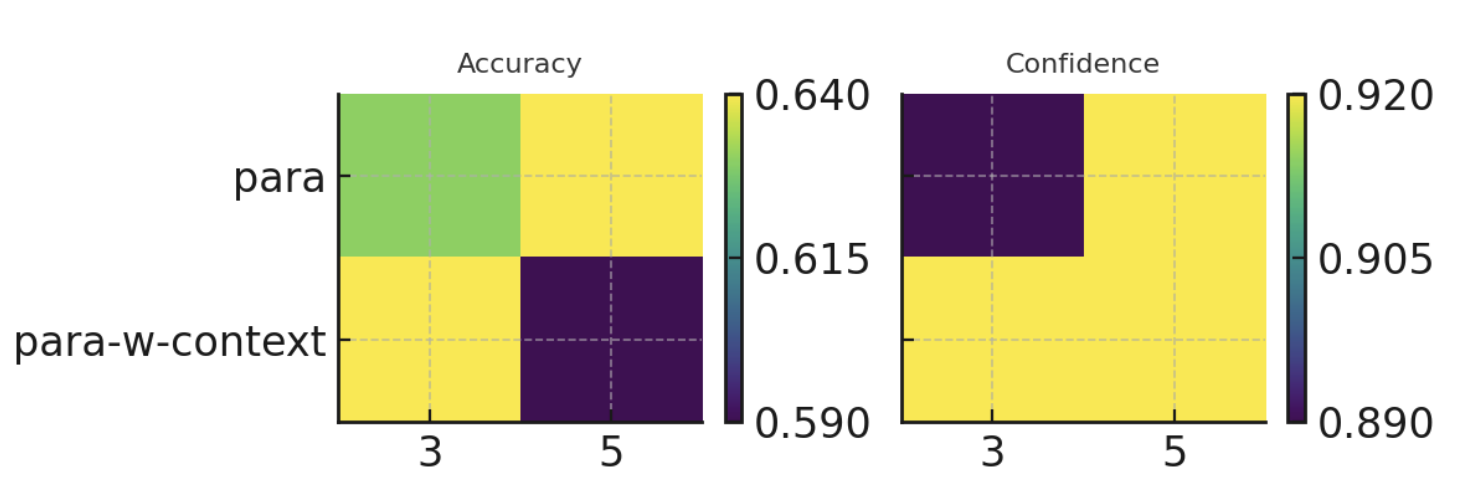}

  \caption{Answer accuracy for paragraph‐only vs.\ paragraph‐with‐context at $k=3,5$.}
  \label{fig:paragraph-chunking}
\end{figure}

To test whether larger semantic units help generation, we evaluated two paragraph‐based schemes at $k\!=\!3$ and $5$:
\begin{itemize}[leftmargin=*,itemsep=2pt]
  \item \emph{Paragraph:} split on natural paragraph boundaries (min. 120 tokens).  
  \item \emph{Paragraph + Context:} each paragraph plus its immediate predecessor and successor.
\end{itemize}
To prevent the truncation‐induced hallucinations we observed with 1600‐char, Top‐5 inputs, we increased Ollama context window to 6000 tokens for these experiments.
Figure~\ref{fig:paragraph-chunking} shows answer accuracy for all paragraph variants.  Adding context does not improve — and even slightly degrades - the accuracy, especially at $k\!=\!5$, confirming that simple paragraph splits suffice.

% removed to fit 3-page strict limit imposed by IPAC
%\subsection{Overall Fine‐Tuning Progression}

%Starting from the Gemma2 baseline (1600‐char, Top‐5, no‐translation), we fine‐tuned on plain paragraphs, then explored all Gemma3 schemes (800/1600 char, Top‐3/5, and the three prompt variants).  Figure~\ref{fig:progress-overal} traces accuracy gains through each step.  The clear winner is 800‐char chunks with Top‐5 retrieval under Gemma3.

%\begin{figure}[htb]
%  \centering
%  \includegraphics[width=\linewidth]{progress.png}
%  \caption{Accuracy progression from Gemma2 baseline through successive Gemma3 fine-tuning schemes.}
%  \label{fig:progress-overal}
%\end{figure}

\textbf{Recommendation.}  
For our accelerator documentation, \emph{800‐char chunks with Top‐5 retrieval} delivers the highest answer accuracy and confidence, illustrating the critical role of chunk size, Top-$k$, and prompt structure. % in domain‐specific RAG.  

%%%%%%%%%%%%%%%%%%%%%%%%%%%%%%%%%%%%%%%
\section{PERSPECTIVES}
% Rasmus
Initial evaluations demonstrate strong performance utilizing a RAG pipeline for question answering based on textual document content.  However, a significant limitation remains in effectively retrieving information embedded within non-textual elements of the source documents, specifically tables and figures encompassing schematics, plots, and photos. 

Current RAG implementations exhibit difficulty in retrieving and reasoning with visual information. Future work will focus on addressing this challenge through two paths: first, exploring pre-processing techniques such as automatic figure captioning to generate descriptive text prior to indexing, thereby enabling textual retrieval of visual content using the same RAG system that we are presently using; and second, the evaluation of multi-modal embedding models that would directly encode figure content into the vector database.

A successful implementation of these enhancements will pave the way to apply this RAG system for other domains.
%, such as electron beam instrumentation. 
%PSI operates two electron accelerators, the Swiss Light Source and SwissFEL. 
We anticipate that our system could significantly accelerate the access to knowledge specific of other PSI facilities
like SLS and SwissFEL.

%%%%%%%%%%%%%%%%%%%%%%%%%%%%%%%%%%%%%%%
\section{ACKNOWLEDGEMENTS}

%The authors would like to express their sincere gratitude to Rudolf Doelling for providing the majority of the documents used in this study. Discussions with Marco Bocchio from Transmutex were particularly inspiring with regard to the graphical user interface. Last but not least, we are deeply thankful to Yingqiang Gao from the University of Zurich for his invaluable technical support.

The authors would like to express their sincere gratitude to Rudolf Doelling, Marco Bocchio and Yingqiang Gao. 
%from the University of Zurich for his invaluable technical support.

%
% only for "biblatex"
%

% end \ifboolexpr
%
% for use as JACoW template the inclusion of the ANNEX parts have been commented out
% to generate the complete documentation please remove the "%" of the next two commands
% 
%%%\newpage

%%%\include{annexes-A4}
\bibliographystyle{unsrt}   % or plain, IEEEtran, etc.
\bibliography{references}

\begin{thebibliography}{15}

\bibitem{Seidel:2010zz}
M.~Seidel, S.~Adam, A.~Adelmann, C.~Baumgarten, R.~Dolling, H.~Fitze,
  A.~Fuchs, J.~Grillenberger, M.~Humbel, D.~Kiselev, \emph{et~al.}
\newblock Production of a 1.3 MW Proton Beam at PSI.
\newblock In \emph{Proceedings of IPAC 2010}, TUYRA03, 2010.

\bibitem{Mayet:2024ypj}
F.~Mayet.
\newblock GAIA: A General AI Assistant for Intelligent Accelerator Operations.
\newblock \emph{arXiv preprint} arXiv:2405.01359, 2024.

\bibitem{Sulc:2024ssg}
A.~Sulc, A.~Bien, A.~Eichler, D.~Ratner, F.~Rehm, F.~Mayet, G.~Hartmann,
  H.~Hoschouer, H.~Tuennermann, J.~Kaiser, \emph{et~al.}
\newblock Towards unlocking insights from logbooks using AI.
\newblock In \emph{Proceedings of IPAC 2024}, THPR37, 2024.
\newblock doi:10.18429/JACoW-IPAC2024-THPR37.
\newblock Also available as \emph{arXiv}:2406.12881 [physics.acc-ph].

\bibitem{Dolling:2003yw}
R.~Dolling.
\newblock Diagnostics of the PROSCAN proton-therapy beam lines.
\newblock In \emph{Proceedings of DIPAC 2003}, 2003.

\bibitem{Sapinski:2022aqv}
M.~Sapinski, R.~D\"{o}lling, and M.~Rohrer.
\newblock Commissioning of the Renewed Long Radial Probe in PSI Ring Cyclotron.
\newblock In \emph{Proceedings of IBIC 2022}, MOP19, 2022.
\newblock doi:10.18429/JACoW-IBIC2022-MOP19.

\bibitem{Forck2011}
P.~Forck.
\newblock JUAS Lecture Notes on Beam Instrumentation and Diagnostics.
\newblock 2011. Available at
  \url{https://www.gsi.de/work/gesamtprojektleitung_fair/commons/beam_instrumentation/research_and_development_rd/veroeffentlichungen.htm}.

\bibitem{gao2023retrieval}
Y.~Gao, Y.~Xiong, X.~Gao, K.~Jia, J.~Pan, Y.~Bi, Y.~Dai, J.~Sun, H.~Wang, and
  H.~Wang.
\newblock Retrieval-Augmented Generation for Large Language Models: A Survey.
\newblock \emph{arXiv preprint} arXiv:2312.10997, 2023.

\bibitem{ollama}
Ollama.
\newblock Ollama project homepage.
\newblock 2024. \url{https://ollama.com/}.

\bibitem{Wang:2024mineru}
B.~Wang, C.~Xu, X.~Zhao, L.~Ouyang, F.~Wu, Z.~Zhao, R.~Xu, K.~Liu, Y.~Qu,
  F.~Shang, B.~Zhang, L.~Wei, Z.~Sui, W.~Li, B.~Shi, Y.~Qiao, D.~Lin, and
  C.~He.
\newblock MinerU: An Open-Source Solution for Precise Document Content
  Extraction.
\newblock \emph{arXiv preprint} arXiv:2409.18839, 2024.

\bibitem{Chen:2024bge}
J.~Chen, S.~Xiao, P.~Zhang, K.~Luo, D.~Lian, and Z.~Liu.
\newblock BGE M3-Embedding: Multi-Lingual, Multi-Functionality,
  Multi-Granularity Text Embeddings Through Self-Knowledge Distillation.
\newblock \emph{arXiv preprint} arXiv:2402.03216, 2024.

\bibitem{Pinecone:2023}
Pinecone.
\newblock Evaluation Measures in Information Retrieval.
\newblock 2023.
\newblock \url{https://www.pinecone.io/learn/offline-evaluation/}.

\bibitem{Ollama_context_window_2024}
Ollama.
\newblock Questions about context size (GitHub issue \#2204).
\newblock 2024. \url{https://github.com/ollama/ollama/issues/2204}.

\bibitem{GemmaTeam2025Gemma3}
Gemma Team.
\newblock Gemma 3.
\newblock 2025. Kaggle. \url{https://goo.gle/Gemma3Report}.

\bibitem{GemmaTeam2024Gemma}
Gemma Team.
\newblock Gemma.
\newblock 2024. Kaggle. doi:10.34740/KAGGLE/M/3301.
\newblock \url{https://www.kaggle.com/m/3301}.

\bibitem{Riviere2024}
M.~Rivi\`{e}re \emph{et~al.}
\newblock Gemma 2: Improving Open Language Models at a Practical Size.
\newblock \emph{arXiv preprint} arXiv:2408.00118, 2024.

\end{thebibliography}

\section{Appendix}

\subsection{Generation Prompt}

\begin{lstlisting}[caption={Prompt without chunk scores},breakautoindent=false,breakindent=0pt]
You are a beam-accelerator Q&A assistant. Answer the user's question using ONLY the provided context. Do NOT add commentary or summarize the context. Match the answer language to the question. If the question requests a numerical value, include the number and unit. If the answer cannot be found in context, respond exactly: "I don't know." Keep the answer to at most two sentences.

QUESTION: <QUESTION>
CONTEXT: <CONTEXT>
\end{lstlisting}

\begin{lstlisting}[caption={Prompt with chunk scores},breakautoindent=false,breakindent=0pt]
You are a beam-accelerator Q&A assistant. Answer the user's question using ONLY the provided context. Do NOT add commentary or summarize the context. Match the answer language to the question. If the question requests a numerical value, include the number and unit. If the answer cannot be found in context, respond exactly: "I don't know." Keep the answer to at most two sentences. Please pay more attention to the higher ranked chunks.

QUESTION: <QUESTION>
CONTEXT with scores: <CONTEXT WITH SCORE>
\end{lstlisting}

\subsection{Evaluation Prompt}

\begin{lstlisting}[caption={Model as Judge}, breakautoindent=false,breakindent=0pt]
You are a strict evaluator. Compare the GENERATED ANSWER to the GOLDEN ANSWER.
- If the generated answer is at least partially correct, respond EXACTLY with:
  {"label": "yes", "confidence": <float 0-1>}
- If it is completely incorrect, respond EXACTLY with:
  {"label": "no",  "confidence": <float 0-1>}
Do NOT output any other text.

QUESTION:
<QUESTION>

GOLDEN ANSWER:
<GOLD>

GENERATED ANSWER:
<GENERATED>
\end{lstlisting}

\end{document}